\documentclass[11pt]{article}
\usepackage{moriond,epsfig}
\usepackage{wrapfig}
\usepackage{amsmath}
\usepackage[usenames]{color}
\usepackage{latexsym}
\usepackage{picinpar}
\usepackage{graphics}
\def\npbps#1#2#3{  { Nucl. Phys. }(Proc. Suppl.){\bf B #1} (19#2) #3}
\def\plb#1#2#3{    { Phys. Lett. }{\bf B #1} (19#2) #3}

\begin{document}
\begin{flushright}
{\small
CERN-TH-2002-213\\
IFUM-713/FT\\
FTUAM-02-422
}
\end{flushright}
\vspace*{4cm}

\title{SOLVING THE SOLAR NEUTRINO PROBLEM WITH KAMLAND AND  BOREXINO}
\author{\underline{P. Aliani}$^{1,2,4,a}$, V. Antonelli$^{1,2,a}$, R. Ferrari$^{1,2,a}$,
  M. Picariello$^{1,2,a}$,
  E. Torrente-Lujan$^{3,4}$ \footnote{paola.aliani@cern.ch,
    vito.antonelli@mi.infn.it, ruggero.ferrari@mi.infn.it,
    marco.picariello@mi.infn.it, emilio.torrente-lujan@cern.ch}}
\address{$^{\text{1}}$ Dip. di Fisica, Universit\`a di Milano,
  $^{\text{2}}$I.N.F.N.,  Sezione di Milano,\\ $^{\text{3}}$ Universidad Autonoma
  de Madrid,$^{\text{4}}$ CERN TH-division, Switzerland}
\maketitle
\abstract{We analyze the expected signals of two future neutrino
  experiments, kamLAND and BOREXINO. We show that with just these experiments,
  we will hopefully be able to determine which of the existing solutions to
  the solar neutrino problem is the \emph{real} solution. We also
  analyze existing solar neutrino data and determine the best-fit
  points in the oscillation-parameter space finding that with the inclusion
  of SNO-charged current, the global-rates analysis gives a favored LMA
  solution with a goodness of fit (g.o.f) of just 32.63 \%, whereas the g.o.f
  of the SMA solution is 9.83\%. Nonetheless,  maximal and quasi-maximal mixing is not favored. If we include the Superkamiokande spectrum in our $\chi^{2}$ analysis, we obtain a LMA solution with a g.o.f. of 84.38 \%.} 
\vskip -0.7truecm
\section{Introduction}\label{introduction}
\vskip -0.3truecm
As the 30$^{\text{th}}$ anniversary of Homestake's first solar neutrino flux
measurement~\cite{Homestake} is coming to pass, there is finally hope that with
the results of just two up-coming experiments we will solve the solar neutrino problem. 
We show that with BOREXINO and kamLAND we will probably be able to identify \emph{the} solution to the solar neutrino problem, which has anguished scientists for the past three decades. We have computed the kamLAND~\cite{kamland} and BOREXINO~\cite{Bonetti:kd} expected signals in order to determine what new information they will be able to provide. As in our previous work~\cite{ours}, we produce iso-signal plots in the solar-neutrino parameter space and confront them with our own exclusion plots for currently-running and past solar neutrino experiments.

It is well known in the neutrino community that with the help of the
charged-current results from the SNO detector, oscillations into
\emph{active} flavors have been confirmed. 
The appearance of a flux of muon
and tau neutrinos from the direction of the sun and their subsequent
charged-current interactions on deuterium not only has strongly disfavored
the pure sterile-neutrino scenario, but has also disfavored one of the
pre-existing solutions to the solar neutrino problem, namely the SMA
solution.
\vskip -0.7truecm
\section{Statistical Analysis and Results}
\vskip -0.3truecm
We describe in limited detail what we obtain from our analysis of existing experiments. We include Homestake~\cite{Homestake}, GallEX/GNO and SAGE~\cite{gallex}, SuperKamiokande~\cite{Nakahata:dt} and SNO-CC (charged-current)~\cite{Poon:2001ee}. We use the BPB2001~\cite{Bahcall:2000nu} neutrino fluxes with the old $S_{17}\left(0\right)$ factor and the 1.25Kdays data for SuperKamiokande. In our simplest analysis, we use the global rates for all the experiments and confront our parameter-dependent signal with the experimental results to determine the best-fit points with the usual $\chi^{2}$ method. We then introduce the SK spectrum and re-determine our local minima.

\begin{center}
\begin{figure}[h]
\vskip -1truecm
\caption{({\sl left}) Global analysis after the inclusion of SNO-CC results. ({\sl right}) analysis including SK spectrum and SNO-CC. Allowed regions at $\chi^{2}$=4.61, 5.99, 9.21 \& 11.83. Black curve represents CHOOZ upper-bound}\label{exclusionPlots}
\begin{tabular}{cc}
\includegraphics[scale=0.6]{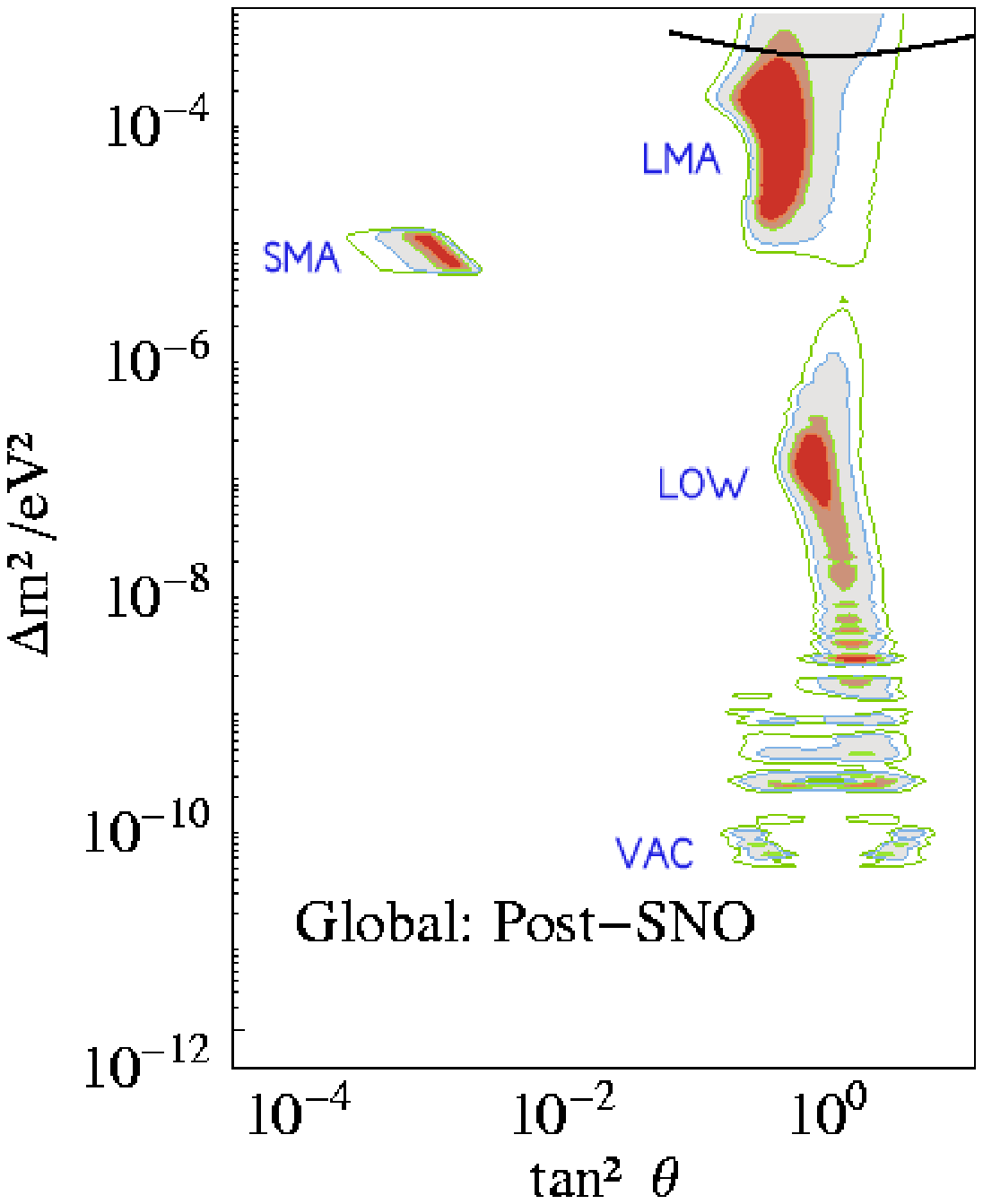}&
\includegraphics[scale=0.6]{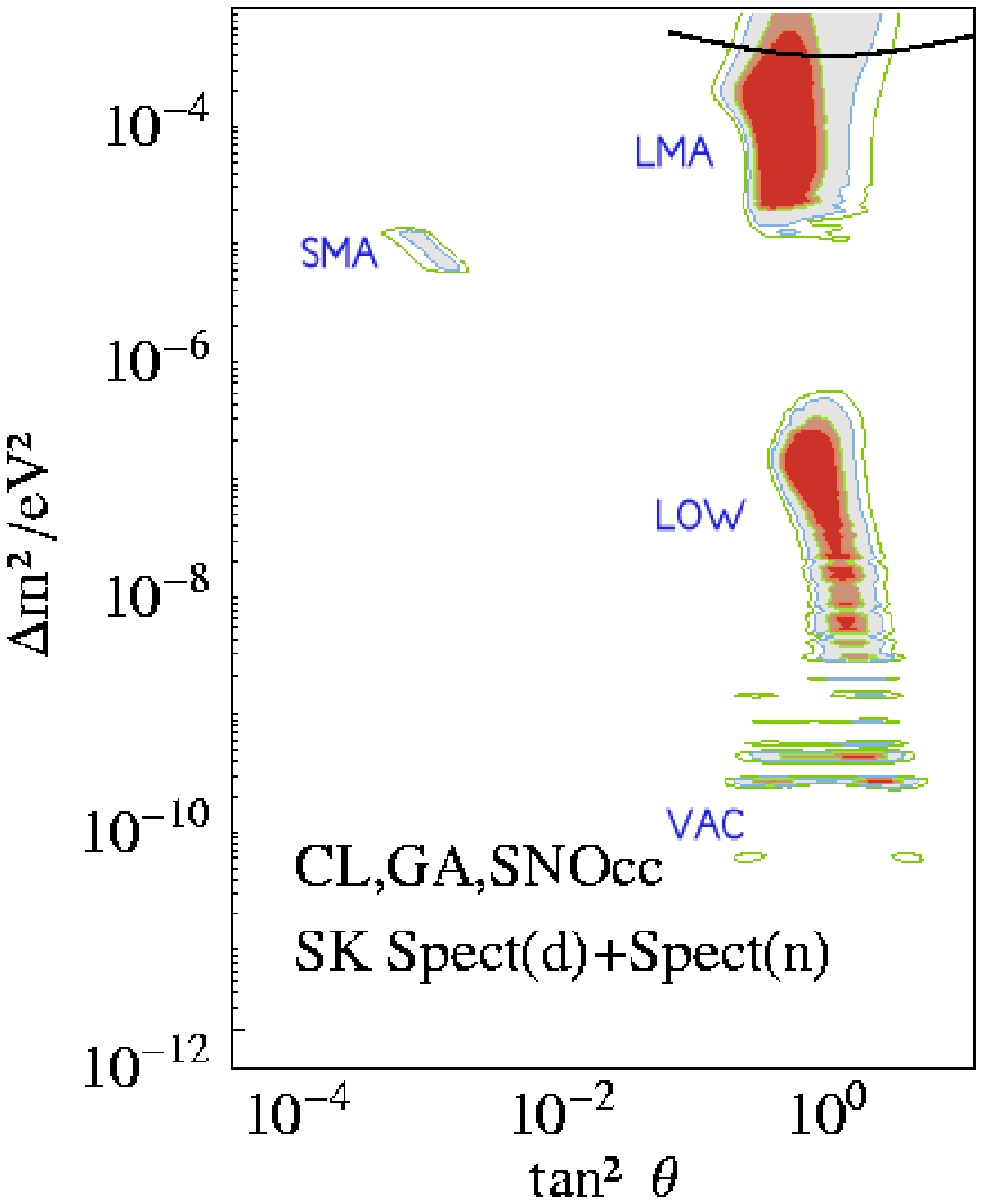}\\
\end{tabular}
\vskip -1.3truecm
\end{figure}
\end{center}

The covariance matrix used is made of two parts: the first is diagonal (for the theoretical, statistical and uncorrelated errors) and the second is off-diagonal (for the correlated systematic uncertainties). We perform a minimization of the $\chi^{2}$ as a function of the oscillation parameters in order to determine the best-fit points. The exclusion plot for the global-rates analysis and the global rates plus SK spectrum are shown in Fig. \ref{exclusionPlots}.\\
As in other similar analyses, four distinct regions in the solar-neutrino parameter space are allowed: the Small Mixing Angle (SMA) solution, the Large Mixing
Angle (LMA) solution, the LOW mass (LOW) solution and the VACuum (VAC)
solution. With the aid of the SNO-CC results, the solutions become well
separated and the SMA solution becomes disfavored over
the LMA. The addition of the SK spectrum in the analysis has the effect of further decreasing the g.o.f. of the SMA solution.
\vskip -0.7truecm
\section{BOREXINO and kamLAND implications}
\vskip -0.3truecm
Although kamLAND is not solely a solar neutrino experiment, it is sensitive to
the currently-favored LMA solution in the solar neutrino parameter space. If
we are lucky, this means that kamLAND results will allow us to determine these  with
high accuracy. Low energy $\overline{\nu}_{e}$
coming from 16 commercial nuclear reactors converge to a scintillator-detector
capable of detecting their interactions with protons via coincidence
measurement of a prompt $\gamma$ emitted from the annihilation of the positron
with the surrounding medium and a delayed $\gamma$ coming from the re-capture
of the emitted neutron~\cite{kamland}. The simulation of the reactor-kamLAND 
detector is
somewhat simpler than that of solar-neutrino detectors due to the fact that
matter effects can be ignored completely. As we mentioned before, the distances
traveled by and the energy of these neutrinos limits the sensitivity of the
detector to the LMA region of the solar-neutrino oscillation-parameter
space. It is known~\cite{Schonert:2002ep} that the sensitivity of kamLAND to the upper
part of the LMA region will not be sufficient to give a good estimate of $\Delta m_{12}^{2}$. The HLMA project must be kept in mind if the case be that the overall $\chi^{2}$ has its minimum at $\Delta m^{2}_{12} > 10^{-3}$ eV$^{2}$. In order to produce our expected kamLAND signal, we have considered a
constant, standard reactor fuel composition, and expected the reactors to run at
around 80\% efficiency~\cite{kamland}.
\begin{center}
\begin{figure}[h]
\vskip -0.85truecm
\caption{BOREXINO expected signals ({\sl left}): Outermost line (\textcolor{Magenta}{magenta}) corresponds to $S_{\text{data}}/S_{\text{SSM}}=0.7$. Following towards the center, (\textcolor{Blue}{blue}), $S_{\text{data}}/S_{\text{SSM}}=0.6$ and in in \textcolor{Red}{red},  $S_{\text{data}}/S_{\text{SSM}}=0.5$. ({\sl right}): day-night asymmetry plot for expected borexino signals in the LMA and LOW-VAC regions. Stars indicate our best-fit points.}\label{expectedsignals}
\begin{tabular}{cc} 
%\hskip -0.4cm
\includegraphics[scale=0.6]{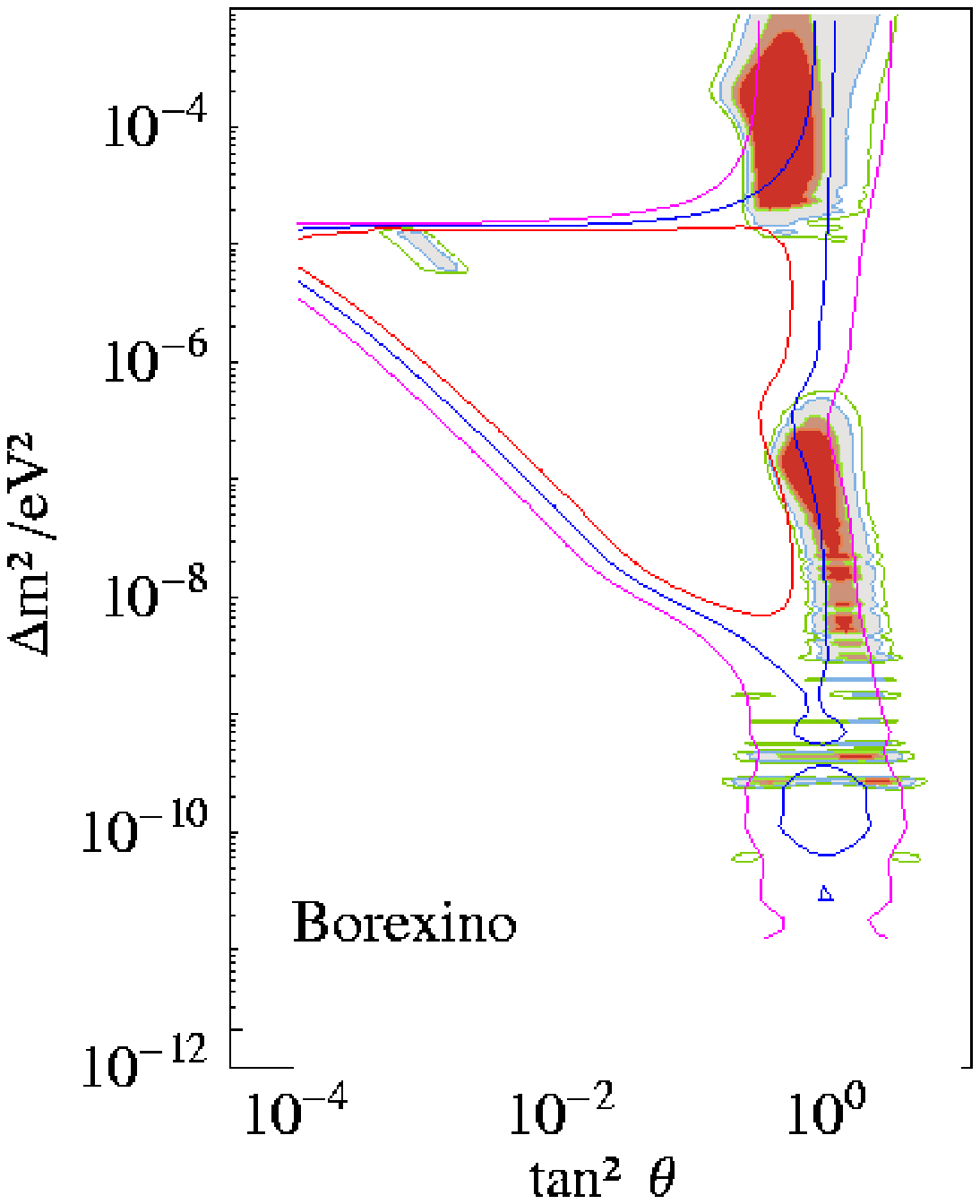}&
%\hskip -0.2truecm
\includegraphics[scale=0.6]{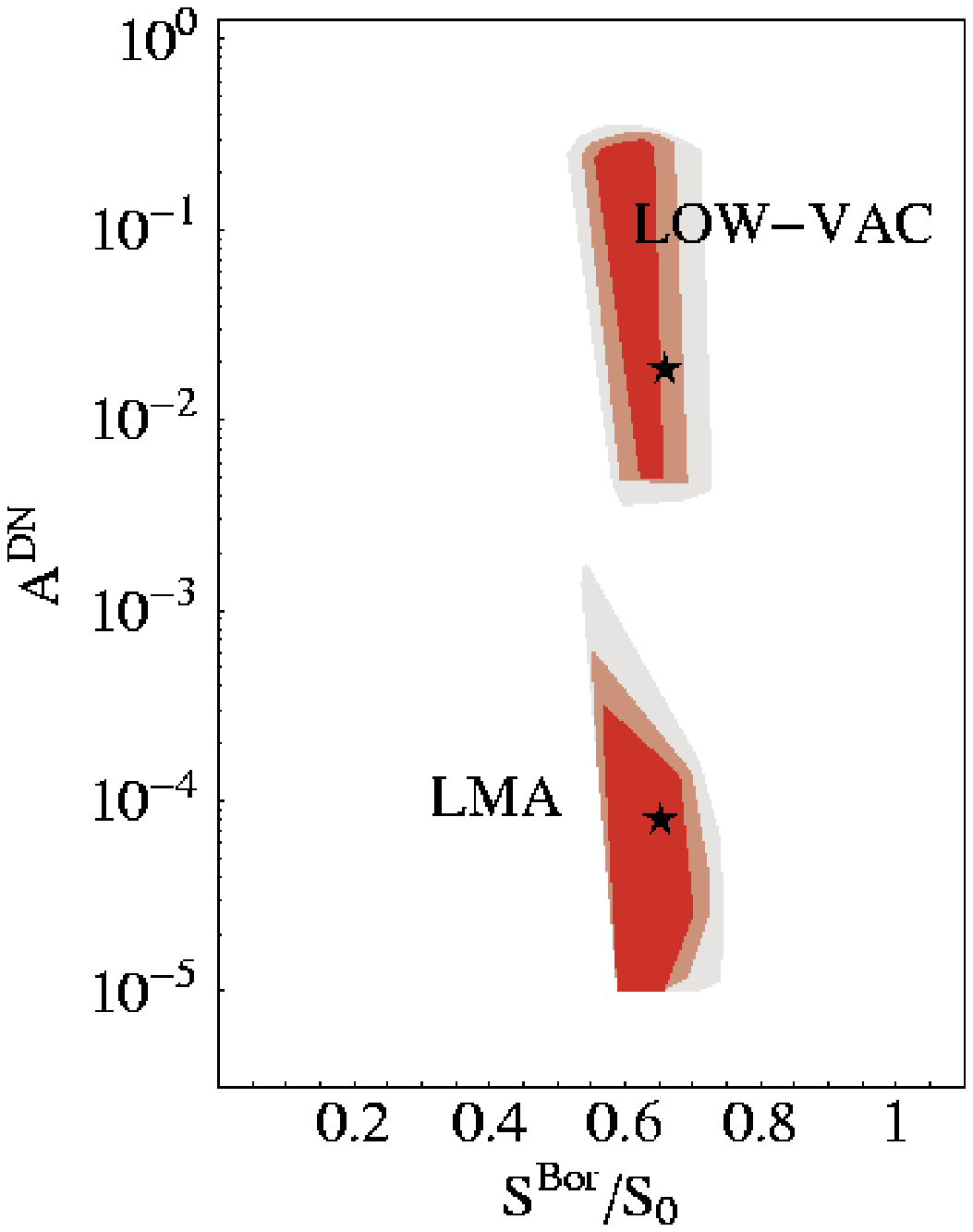}\\
\end{tabular}
\vskip -1.5truecm
\end{figure}
\end{center}
BOREXINO is a solar-neutrino experiment, whose detector is similar to
kamLAND. It will be sensitive primarily to the $^{7}$Be neutrinos although the
collaboration has aims to measure solar anti-neutrinos as well. In simulating
the BOREXINO expected signal, solar and earth matter effects cannot be
neglected. We have therefore also calculated the day and night expected signals and the expected day-night asymmetry.
\begin{figure}[ht]
\label{asymmetry}
\vskip -0.75truecm
\caption{kamLAND expected signal. Innermost line (black), $S_{\text{data}}/S_{\text{No osc.}}=0.4$, following, $S_{\text{data}}/S_{\text{No osc.}}=0.7$ (\textcolor{Blue}{blue}), $S_{\text{data}}/S_{\text{No osc.}}=0.8$ (\textcolor{Red}{red}) and  $S_{\text{data}}/S_{\text{No osc.}}=0.95$ (\textcolor{magenta}{magenta}).}
\begin{center}
\includegraphics[scale=0.6]{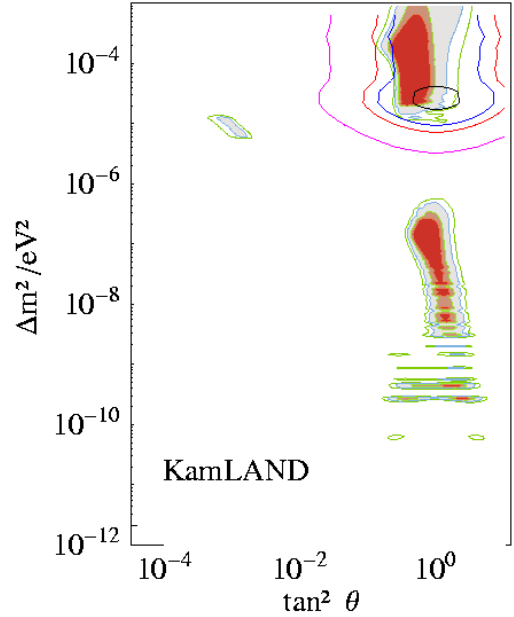}
\end{center}
\vskip -0.5truecm
\end{figure}
It is clear that as the kamLAND expected signal decreases, (Fig. \ref{asymmetry}) a smaller and smaller portion of the LMA solution is compatible with precedent results, whereas the BOREXINO signal is capable only of distinguishing between the LMA-LOW-VAC solutions and the SMA solution by the discriminating $S_{\text{data}}/S_{\text{SSM}}=0.5$ line. If we combine these two results with the analysis of the expected day-night asymmetry $\mathcal{A}=2\left(D-N\right)/\left(D+N\right)$ presented in Fig. \ref{expectedsignals} (right panel) we see that BOREXINO alone will distinguish among the three solutions. If the day-night asymmetry is negligible, the solution lies in the LMA region. If the asymmetry is $> 5\times 10^{-3}$ then the solution must lie in the LOW-VAC region.
\vskip -0.7truecm
\section{Conclusions}
\vskip -0.3truecm
We have shown that with the inclusion of just two new experiments, the solution to the solar neutrino problem will be determined. If we are lucky, this solution will lie in the now-favored LMA region and we will benefit from  kamLAND's potentiality for determining with high accuracy just what the mixing parameters are. If BOREXINO excludes the large-mixing-angle solutions, we will still determine with unprecedent precision the mixing parameters due to the already-small extension of the SMA solution \footnote{\textcolor{Sepia}{At ``Les Rencontres de Moriond 2002'', SNO-NC results had not yet been published. SNO has again recently proved its worth~\cite{snoNew}. The new data confirms the LMA solution~\cite{recentSNOnc}, which is rather fortunate for the kamLAND collaboration.}}.

\end{document}